\documentclass[Times,8pt,aps,prl,onecolumn,showpacs,amsmath,amssymb,floatfix,footinbib,superscriptaddress]{revtex4-1}
\usepackage{amsmath,natbib,graphicx,subfigure,amssymb,graphics,amsmath,mathrsfs,CJK,color}
\usepackage{multirow,fancyhdr,color,bm,tabularx,psfrag,geometry,dcolumn,datetime}
\usepackage[colorlinks=true,linkcolor=blue,urlcolor=blue,citecolor=blue]{hyperref}
\usepackage[mathlines]{lineno}
\def\footnoterule{\kern -1mm \hrule width 5.8cm \kern 2.2mm}%
\usepackage{tikz,xcolor,hyperref}
\definecolor{lime}{HTML}{A6CE39}
\DeclareRobustCommand{\orcidicon}{%
 \begin{tikzpicture}
 \draw[lime, fill=lime] (0,0)
    circle [radius=0.16]
    node[white] {{\fontfamily{qag}\selectfont \tiny ID}};\draw[white, fill=white] (-0.0625,0.095)
    circle [radius=0.007];
 \end{tikzpicture}
\hspace{-2mm}}
\foreach \x in {A, ..., Z}
{\expandafter\xdef\csname orcid\x\endcsname{\noexpand\href{https://orcid.org/\csname orcidauthor\x\endcsname}{\noexpand\orcidicon}}}

\begin{document}


\title{ Localization of cold $^{87}Rb$ atom within half-wavelength domain }


\author{Shun-Cai Zhao\orcidA{}}
\email[Corresponding author: ]{ zhaosc@kmust.edu.cn }
\affiliation{Department of Physics, Faculty of Science, Kunming University of Science and Technology, Kunming, 650500, PR China}

\author{Xin Li}
\affiliation{Department of Physics, Faculty of Science, Kunming University of Science and Technology, Kunming, 650500, PR China}

\author{Ping Yang}
\affiliation{Department of Physics, Faculty of Science, Kunming University of Science and Technology, Kunming, 650500, PR China}

\date{\today}

\begin{abstract}
Simulating the cold $^{87}Rb$ atom with a three-level quantum system interacting with two orthogonal standing-wave fields,
 the localization within half-wavelength domain in the x-y plane is achieved by monitoring the probe absorption.
 Within the half-wavelength domain, the single absorption peak increases from 0.2 to 1.0 via the spontaneously generated coherence (SGC),
 while the diameters of the single absorption peaks are diminished by the increasing incoherent pumping field. Our scheme provides
 the flexible parameters manipulating manner for the localization of cold $^{87}Rb$ atom.
\end{abstract}
\pacs{  }


\maketitle
Cold atoms are widely utilized in the precise measurement of their fundamental physical constants\cite{1,2}
and their internal structures. And not only that, they make the promising resource in quantum information
for generation, manipulation, and storage of quantum state of light\cite{3,4}. Therefore, localizing the
neutral cold atoms has been an important standard technique in atomic optics and quantum optics.
In the search of hybrid systems consisting of isolated atoms and solid devices,
the key character is the ability to localize, manipulate cold atoms near the surface of the solid device in
a nanoscale region. Recently, several schemes for localizing and manipulating cold
atoms in a nanoscale region have been proposed, for instance, diffraction by structures with sub-wavelength
characteristic sizes\cite{5,6,7,8}, nanometer metallic plasmonic devices\cite{9,10,11,12,13,14} and one-dimensional light
localization of the evanescent light wave\cite{15,16}.

In this paper, we theoretically investigate localization for cold $^{87}Rb$ atom
via the simulation of a three-level quantum system interacting with two orthogonal standing-wave field
and an incoherent pumping field. The single localization peak monitored by the probe absorption is obtained
within 0.5 wavelength domain. The single localization peak with its diameter being smaller than the Rayleigh
limit of 0.5 wavelength demonstrates a precise 2D atom localization for cold $^{87}Rb$ atom.
What's more, a very important advantage in our scheme is that we provide a realistic cold $^{87}Rb$ atomic system
for realizing 2D atom localization by monitoring the probe absorption, which may
make our scheme much more convenient in experimental realization.

\section{MODEL AND EQUATION}
A proposed scheme to generate the localized cold atoms $^{87}Rb$ is shown in Fig. 1.
The cold atoms $^{87}Rb$ was simulated by a $\Lambda$-type three-level system with an
upper energy level $|1\rangle$ and two closely lying lower energy levels $|2\rangle$
and $|3\rangle$. In the sample, that the $5S_{1/2}$-$5P_{3/2}$ transition is assumed to be the $D_{2}$ line \cite{17}.
The levels of $|5S_{3/2},F$=$2\rangle$, $|5S_{1/2},F$=$1\rangle$ and $|5S_{1/2}$,$F$=$2\rangle$
in the cold atoms $^{87}Rb$ behave the $|1\rangle$, $|2\rangle$ and $|3\rangle$ states, respectively.
A weak probe field $E_{p}$ of frequency $\nu_{p}$ with Rabi frequencies
$\Omega_{p}$ and an incoherent pump process represented by a
rate 2$\Gamma$ simultaneously drive the transition $|1\rangle$ and $|2\rangle$.
The standing-wave field $E_{x,y}$ of frequency $\nu_{c}$ with the superposition
of two orthogonal standing-wave fields, i.e., one is in the x
direction and the second is along y direction coupling the levels $|1\rangle$
and $|3\rangle$. The Rabi frequency corresponding to the probe field $E_{p}$ is
$\Omega_{p}$=$E_{p}\mu_{13}/2\hbar$, and the position dependent Rabi frequency
of the standing-wave field $E_{x,y}$ is $\Omega_{c}(x,y)$=$E_{x,y}\mu_{12}/2\hbar$,
where $\mu_{13}$ and $\mu_{12}$ are the corresponding dipole
matrix elements. 2$\gamma_{1}$ and 2$\gamma_{2}$ are the spontaneous emission rates
from level $|1\rangle$ to levels $|3\rangle$ and $|2\rangle$, respectively.
It should be noted that the SGC effect is remarkable when the energy spacing between
the two lower levels is small\cite{18}. When the energy spacing is large, the rapid oscillation between $|2\rangle$ and
$|3\rangle$ will average out such effect.

\vskip 2mm
\begin{figure}[htp]
\includegraphics[width=0.45\columnwidth]{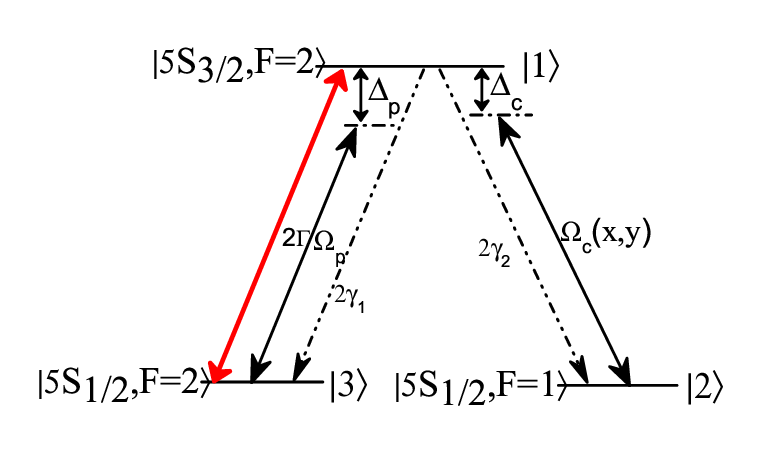}\includegraphics[width=0.45\columnwidth]{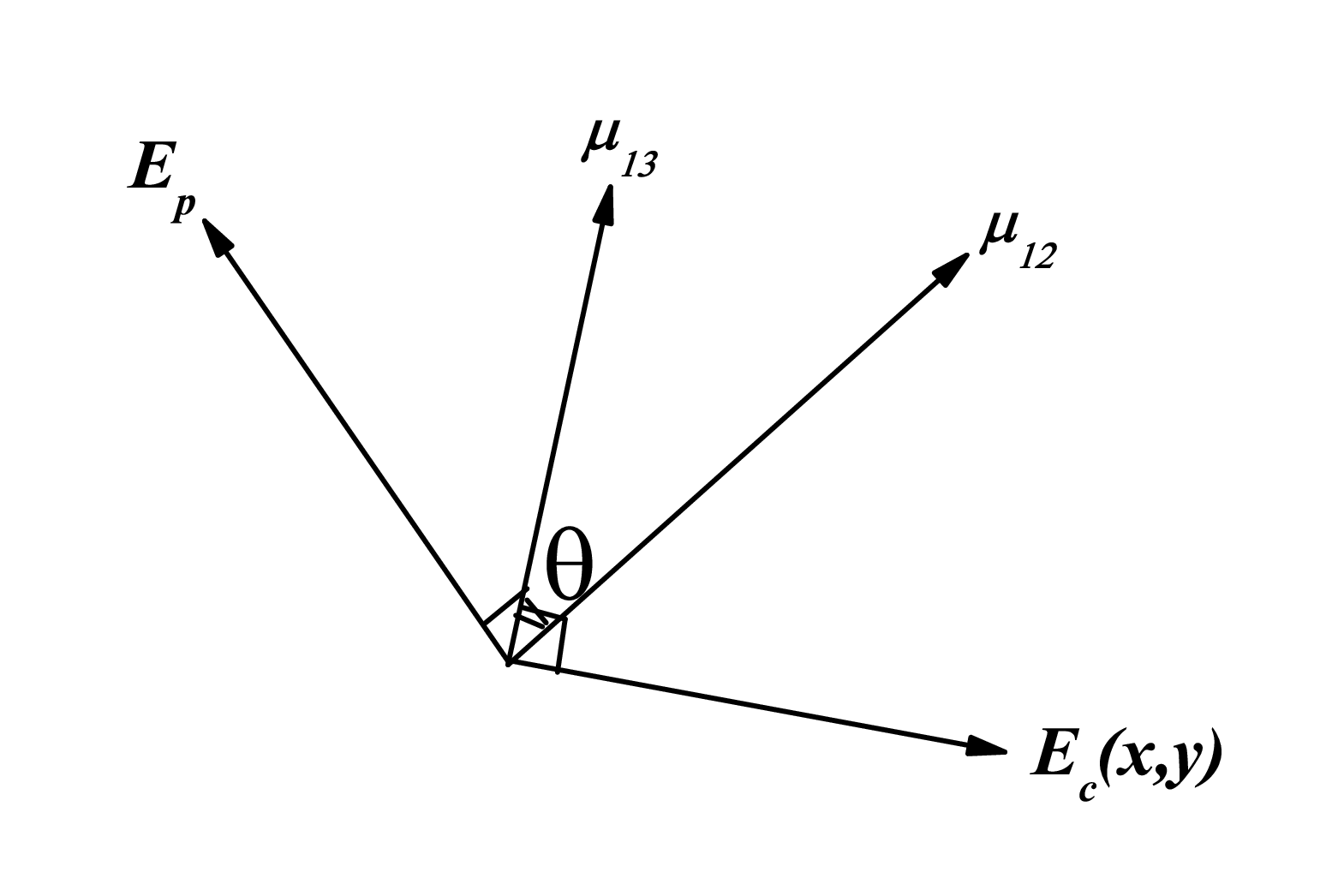}
\caption{(Color online)Schematic diagram of a $\Lambda$-type atomic system for the cold atom $^{87}Rb$: the levels of $|5S_{3/2},F$=$2\rangle$, $|5S_{1/2},F$=$1\rangle$ and $|5S_{1/2}$,$F$=$2\rangle$ in the cold atoms $^{87}Rb$ behaving the $|1\rangle$, $|2\rangle$ and $|3\rangle$ states, respectively. A standing-wave field with position-dependent Rabi frequency $\Omega_{c}(x,y)$
couples the transition $|1\rangle$-$|2\rangle$. A weak probe field and an
incoherent pump field drive the transition $|1\rangle$-$|3\rangle$, simultaneously.
The field polarizations are chosen so that one field drives only one transition.
2$\gamma_{1}$ and 2$\gamma_{2}$ are the decay rates.
}
\end{figure}
\vskip 4mm

The position-dependent Rabi frequency $\Omega_{c}(x,y)$ of the field $E_{c}(x,y)$ superposed by the two standing-wave fields
$E_{c}(x)$ and $E_{c}(y)$ \cite{19,20} is given by

\begin{align}
\Omega_{c}(x,y)=\Omega_{c0}[sin(\kappa_{1}x+\delta)+sin(\kappa_{2}y+\eta)]
\end{align}
where $\kappa_{i}$=$2\pi/\lambda_{i},(i$=$1,2)$ is the wave vector with
wavelengths $\lambda_{i},(i=1,2)$ of the corresponding standing-wave
fields. The parameters $\delta$ and $\eta$ are the phase shifts
associated with the standing-wave fields. The
center-of-mass position of the atom along the direction of the
standing-wave field is assumed to be constant. Therefore, we neglected the
kinetic-energy part of the Hamiltonian under the Raman-Nath
approximation\cite{21}. In the interaction picture and rotating-wave frame, Hamiltonian
of this $\Lambda$ atom system is defined as

\begin{align}
H_{I}&=\Delta_{p}|1\rangle\langle1|+(\Delta_{p}-\Delta_{c})|2\rangle\langle2|-[\Omega_{c}|1\rangle\langle2|\\
      &+\Omega_{p}|1\rangle\langle3|+c.c.].
\end{align}
Here, $\Delta_{p}$=$\omega_{13}-\nu_{p}$ and $\Delta_{c}$=$\omega_{12}-\nu_{c}$
are the field detunings corresponding to the atomic transitions $|1\rangle$-$|3\rangle$
and $|1\rangle$-$|2\rangle$, respectively. $\nu_{c}$ and $\nu_{p}$ are the
frequencies for the coupling standing-wave fields and the weak probe field.

Then using the Weisskopf-Wigner theory of spontaneous emission\cite{22} under
the rotating-wave approximation and the dipole approximation, we get the
density-matrix equations as follows:

\begin{align}
\dot{\rho_{22}}&=2\gamma_{2}\rho_{11}+i\Omega_{c}(\rho_{12}-\rho_{21}),\nonumber\\
\dot{\rho_{33}}&=2\gamma_{1}\rho_{11}-2\Gamma\rho_{33}+i\Omega_{p}(\rho_{13}-\rho_{31}),\tag3\\
\dot{\rho_{12}}&=-(\gamma_{1}+\gamma_{2}+i\Delta_{c})\rho_{12}+i\Omega_{p}\rho_{32}-i\Omega_{c}(\rho_{11}-\rho_{22}),\nonumber\\
\dot{\rho_{13}}&=-(\gamma_{1}+\gamma_{2}+\Gamma+i\Delta_{p})\rho_{13}+i\Omega_{c}\rho_{23}-i\Omega_{p}(\rho_{11}-\rho_{33}),\nonumber\\
\dot{\rho_{23}}&=-(\Gamma+i\Delta_{p}-i\Delta_{c})\rho_{23}+2p\sqrt{\gamma_{1}\gamma_{2}}\rho_{11}+i\Omega_{c}\rho_{13}\nonumber\\
               &-i\Omega_{p}\rho_{21}\nonumber
 \end{align}
The above equations are constrained by $\rho_{11}+\rho_{22}+\rho_{33}=1$ and $\rho_{ij}^{\ast}=\rho_{ji}$.
The effect of SGC is very sensitive to the orientations of the atomic dipole moments $\mu_{13}$ and $\mu_{12}$ \cite{18}.
Here, the parameter p depicts the angle between the two induced dipole moments and is defined as
$p=\mu_{13}\cdot\mu_{12}/|\mu_{13}\cdot \mu_{12}|$=$cos\theta$
with $\theta$ being the angle between the two dipole moments(Seen in Fig.1).
Between 0 and 2$\pi$, $\theta$ can be a random angle except for 0 and $\pi$,
and which is very sensitive for the existence of the SGC effect.
The terms with $p\sqrt{\gamma_{1}\gamma_{2}}$ represent the quantum
interference resulting from the cross coupling between spontaneous emission
paths $|1\rangle$$\leftrightarrow$$|2\rangle$ and $|1\rangle$$\leftrightarrow$$|3\rangle$.

Our goal here is to obtain the 2D atom localization for cold $^{87}Rb$ atom from
the susceptibility of the system at the probe frequency\cite{21}.
The cold $^{87}Rb$ atomic nonlinear Raman susceptibility $\chi$ is then given by

\begin{align}
\chi=\frac{2N|\mu_{13}|^{2}}{\epsilon_{0}\Omega_{p}\hbar}\rho_{13},\tag4\\
\end{align}
where N is the cold $^{87}Rb$ atom number density and $\mu_{13}$ is the magnitude
of the dipole-matrix element between  $|1\rangle$ and $|3\rangle$. $\epsilon_{0}$
is the permittivity in free space. For simplicity we assume $\Omega_{p}$ and $\Omega_{0}$ to be real.
In the limit of a weak probe, the steady-state solutions for $\rho_{13}$
to the first order of $\Omega_{p}$ can be written as
\begin{align*}
\rho^{(1)}_{13}&=\frac{\Omega_{c} \Omega_{p} \rho^{(0)}_{21} - i \Omega_{p} C_{3} \rho^{(0)}_{23}}{\Omega_{c}^{2}-C_{1} C_{3}} + \frac{p\sqrt{\gamma_{1} \gamma_{2}} \Omega_{c}^{2}B_{0}}{(\Omega_{c}^{2}-C_{1} C_{3})B_{1}},\\
\end{align*}
 with
\begin{align*}
\rho^{(0)}_{21}&=\frac{[A_{1}- i (A_{2} + A_{4})] A_{3}}{A_{1}^{2} + A_{2}^{2}- A_{4}^{2}},\\
\rho^{(0)}_{31}&=\frac{2 p \Gamma \Omega_{c}^{2}(-i\Omega_{c}+C_{7}\rho^{(0)}_{21})\sqrt{\gamma_{1}\gamma_{2}}}
                {(\Omega_{c}^{2}+C_{5}C_{6})(\Omega_{c}^{2} C_{4}-2\gamma_{2} \Gamma C_{7})},\\
\rho^{(0)}_{32}&=- \frac{ 2 p \sqrt{\gamma_{1} \gamma_{2}} \Gamma \Omega_{c} C_{1} B_{2} }{(A_{1}^{2} + A_{2}^{2} -
               A_{4}^{2})(\Omega_{c}^{2} - C_{1} C_{3}) B_{3} },\\
\rho^{(0)}_{11}-\rho^{(0)}_{33}&=\frac{\Omega_{c} (\Omega_{c} + C_{2} \rho^{(0)}_{21})(\Gamma - \gamma_{1})}{2 \Gamma (\Omega_{c}^{2} + A_{5})+\gamma_{1} (\Omega_{c}^{2}+2\Gamma\gamma_{2})},\\
\rho^{(1)}_{21}&=- i\frac{2 \Gamma \Omega_{p}  \rho^{(0)}_{23}(C_{9}+\gamma_{1}\gamma_{2})+{\Omega_{c} A_{9} \gamma_{2} B_{4}+ B_{5}}}{2\Omega_{c}^{2} A_{0} C_{4}+2 \Gamma (A_{6}+C_{0}\gamma_{1}) \gamma_{2}},
\end{align*}
\noindent where
\(A_{0}=\gamma_{1}+ \gamma_{2}, A_{1}=2\Gamma\Delta_{c}\gamma_{2},
A_{2}=2\Gamma\gamma_{1}\gamma_{2}+2\Gamma\Omega_{c}^{2}+2\Gamma \gamma_{2}^{2} +\Omega_{c}^{2}\gamma_{1},
A_{3}=2\Gamma\Omega_{c}\gamma_{2},
A_{4}=-2\Gamma\Omega_{c}^{2}-\gamma_{1}\Omega_{c}^{2},
A_{5}=i\gamma_{2}\Delta_{c}+\gamma_{2}^{2},
A_{6}=\Delta_{c}^{2}+\gamma_{2}^{2},
A_{7}=\Delta_{c}+i\gamma_{1}+i\gamma_{2},
A_{8}=\Gamma+i\Delta_{p}+\gamma_{1}+\gamma_{2},
A_{9}=\Delta_{c}-i\gamma_{1}-i\gamma_{2},
B_{0}=[\Omega_{p}\Omega_{c}(\rho^{(0)}_{13}-\rho^{(0)}_{31})+2 i\Gamma(\Omega_{c}+C_{2}\rho^{(1)}_{21}+\Omega_{p}\rho^{(0)}_{32})],
B_{1}=\Omega_{c}^{2} C_{4}-2i\Gamma C_{2}\gamma_{2},
B_{2}=\Omega_{c}( A_{1}^{2}+A_{2}^{2}-A_{4}^{2})+A_{3}C_{2}[ A_{1}+i (A_{2}+A_{4})],
B_{3}=\Omega_{c}^{2} C_{4}+2 i C_{2} \Gamma \gamma_{2},
B_{4}=2i\Gamma+\Omega_{p}(\rho^{(0)}_{13}-\rho^{(0)}_{31}),
B_{5}=\Omega_{c}^{2}\Omega_{p} C_{4} (\rho^{(0)}_{23}-\rho^{(0)}_{32})\),
and
\(C_{0}=\gamma_{1}+2\gamma_{2},
C_{1}=-A_{8}, C_{2}=-\Delta_{c}+i\gamma_{1}+i\gamma_{2},
C_{3}=\Gamma-i(\Delta_{c}-\Delta_{p}),
C_{4}=2\Gamma+\gamma_{1},
C_{5}=-\Gamma-i\Delta_{c}+i\Delta_{p},
C_{6}=-\Gamma-\gamma_{1}-\gamma_{2}+i\Delta_{p},
C_{7}=i\Delta_{c}-\gamma_{1}-\gamma_{2},
C_{8}=-i\Delta_{c}\gamma_{2}+\gamma_{2}^{2},
C_{9}=\gamma_{2}^{2}+i\Delta_{c}\gamma_{2}\).
We have set \(\gamma_{1}=\gamma_{2}=\gamma\),
and all the other parameters are reduced to dimensionless units by scaling with $\gamma $.

Thus the  cold $^{87}Rb$ atomic nonlinear Raman susceptibility
$\chi$ at the probe frequency can therefore be calculated
via Eq.(4), which consists of both real and imaginary parts,
i.e., $\chi=\chi^{'} + i\chi^{''}$. The imaginary part of the susceptibility depicting the
absorption profile of the probe field can be abbreviated as
\vskip -0.6cm
\begin{align}
\chi^{''}
=\alpha Im[\frac{\rho_{13}}{\Omega_{P}}],\tag5
\end{align}
\noindent where $\alpha=\frac{2N|\mu_{13}|^{2}}{\epsilon_{0}\hbar}$. Here we are interested
in the cold atoms $^{87}Rb$ 2D localization via the absorptive process of the
probe field\cite{23}, and Eq.(5) is the main result which can be determined by the
intensities of SGC and the incoherent pump field.

\section{RESULTS AND DISCUSSION}

However, the analytical expression for Eq.(5) corresponding to p depicting the intensities of SGC and to the
incoherent field is rather cumbersome to obtain, so, we follow the numerical approach to
analyze the cold atoms $^{87}Rb$ 2D localization. The single peak with its diameter being smaller than the Rayleigh limit of
0.5 wavelength will demonstrate a precise localization via the SGC and the incoherence pump field.

\begin{figure}[htp]
\center
\includegraphics[width=0.45\columnwidth]{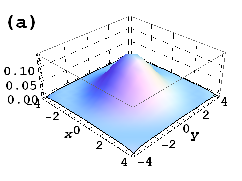}\includegraphics[width=0.45\columnwidth]{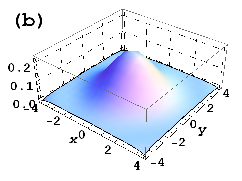}
\includegraphics[width=0.45\columnwidth]{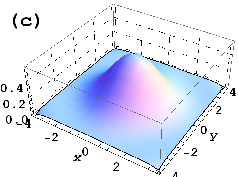}\includegraphics[width=0.45\columnwidth]{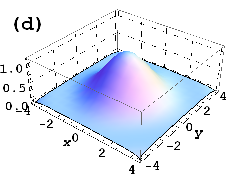}
\caption{$\mathbf{Fig.2}$ (Color online) Plots for the cold atoms $^{87}Rb$ localization: $\chi^{''}$ as a function of
(x,y) in dependence on the intensities of SGC with (a) $\theta$=$\pi/12$,
(b) $\theta$=$\pi/10$, (c) $\theta$=$\pi/7$, (d) $\theta$=$\pi/5$.
$\Gamma$=0.6$\gamma$, $\Delta_{c}$=-10$\gamma$, $\Delta_{p}$=0,
$\delta$=$\eta$=$\pi/2$, $\kappa_{1}$=$\kappa_{2}$=$\pi/6$,
$\Omega_{0}$=2.5$\gamma$, $\Omega_{p}$=0.01$\gamma$, $\gamma_{1}$=$\gamma_{2}$=$\gamma$, where $\gamma$ is the scaling parameter.}
\end{figure}\label{Fig.2}

\begin{figure}[htp]
\center
\includegraphics[width=0.45\columnwidth]{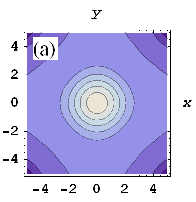}\includegraphics[width=0.45\columnwidth]{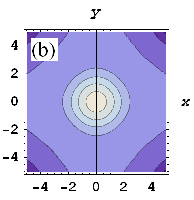}
\includegraphics[width=0.45\columnwidth]{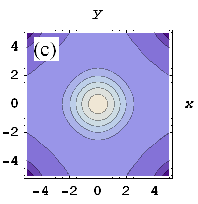}\includegraphics[width=0.45\columnwidth]{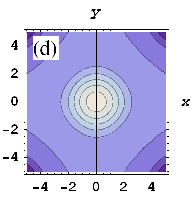}
\caption{$\mathbf{Fig.3}$ (Color online) Plots for the cold atoms $^{87}Rb$ localization: contour plots of $\chi^{''}$ as a function of (x,y) in dependence on the intensities of SGC. All the parameters from (a) to (d) are the same as in Figs.2 from (a) to (d), respectively.}
\end{figure}\label{Fig.3}

Firstly, we set the intensities of incoherent pump field as $\Gamma$=$0.6\gamma$, the wavelengths' parameters $\kappa_{1}$$=\kappa_{2}$$=\pi/6$ which guarantees the wavelengths of the two orthogonal standing-wave fields are 12 wavelength units. And the Rabi frequencies, $\Omega_{c0}$$=$$2.5\gamma$, $\Omega_{p}$$=$$0.01\gamma$.
$\Delta_{c}$$=-10$$\gamma$, $\Delta_{p}$=0$\gamma$. For these
choices of parameters, we consider the cold atoms $^{87}Rb$ localization dependent p within 4 wavelength units in the x-y plane. The intensities of SGC relates p=$cos\theta$ with $\theta$ being set as (a)$\theta$=$\pi/12$,
(b)$\theta$=$\pi/10$, (c)$\theta$=$\pi/7$, (d)$\theta$=$\pi/5$ in Fig.2, respectively.

In Fig.2, we noted that a single localization peak within
the half-wavelength is observed. As it's observed, the peak maxima are about 0.10 in Fig.2(a), 0.20 in Fig.2(b), 0.4
in Fig.2(c), and 1.0 in Fig.2(d). The increasing tendency of peak maxima means a positive
effect on the localization-dependent p, i.e. SGC. In order to get a deeper insight into the effect of p on the cold
atoms $^{87}Rb$ 2D localization, we show their corresponding contour plots in Fig.3.
As shown from Fig.3(a) to Fig.3(d), the diameters of the innermost contour lines
corresponding to the peak maxima are almost the same and are much less than 4 wavelength units, i.e., within
0.5 wavelength, which is much smaller than the Rayleigh limit of 0.5 wavelength.
These demonstrated the intensities of SGC can achieve a precise localization for the cold atoms $^{87}Rb$.

As the above observed, we got a clear portrait for the SGC
in the cold atoms $^{87}Rb$ localization. In the following,
we consider the behavior of the incoherent pump field in  the cold atoms $^{87}Rb$ localization. In Fig.4, we plot the $\chi^{''}$ with $\theta$=$\pi/12$
versus positions x-y for different values of the incoherent pump field, i.e., $\Gamma$ is equal to (a)2.5$\gamma$, (b)4.0$\gamma$, (c)12$\gamma$, (d)15$\gamma$.
And all the other parameters are the same as those in Figs.2. As it notes that the single localizing peak is remained in the x-y plane.

\begin{figure}[htp]
\center
\includegraphics[width=0.45\columnwidth]{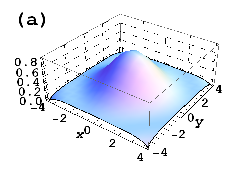}\includegraphics[width=0.45\columnwidth]{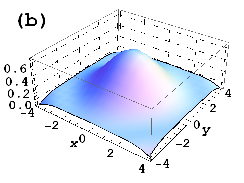}
\includegraphics[width=0.45\columnwidth]{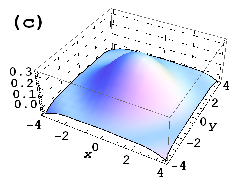}\includegraphics[width=0.45\columnwidth]{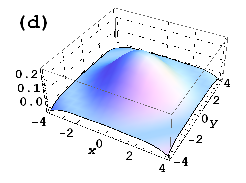}
\caption{$\mathbf{Fig.4}$ (Color online)  Plots for the cold atoms $^{87}Rb$ localization: $\chi^{''}$ as a function of
(x,y) in dependence on the incoherence pump field $\Gamma$. (a)$\Gamma$=2.5$\gamma$,
(b)$\Gamma$=4$\gamma$, (c)$\Gamma$=12$\gamma$, (d)$\Gamma$=15$\gamma$. p=$cos\theta$
with $\theta$ =$\pi/5$. All the other parameters are the same as in Figs.2.}
\end{figure}\label{Fig.4}

However, the single peak maxima are decreasing, i.e. 0.8, 0.6, 0.3 and 0.2 in Fig.4 from (a) to (d).
In a sense, the strong incoherent pumping field plays a destructive role in implementing the
localization for cold atoms $^{87}Rb$. But their corresponding contour diagrams in Fig.5 show
the diameters of the innermost contour lines are shrinking and less than 4 wavelength units
(far less than the Rayleigh limit of 0.5 wavelength). The reason may come from that the incoherent
pump field and weak probe field drive the same transition $|1\rangle$ and $|2\rangle$ simultaneously,
and the increasing incoherent pump field destroys the coherence and deteriorates the probe absorption.
The shrinking diameter of the peak maxima increases the precision for localization.
So, compared to the SGC, an appropriate parameters of the incoherent pump
field will play an important role in the cold atoms $^{87}Rb$ localization.

\begin{figure}[htp]
\center
\includegraphics[width=0.45\columnwidth]{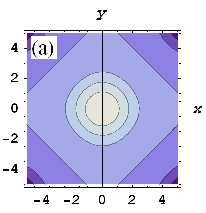}\includegraphics[width=0.45\columnwidth]{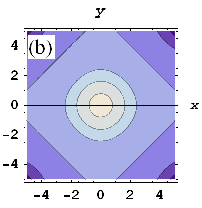}
\includegraphics[width=0.45\columnwidth]{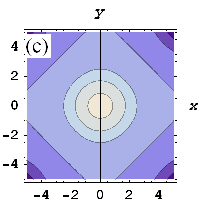}\includegraphics[width=0.45\columnwidth]{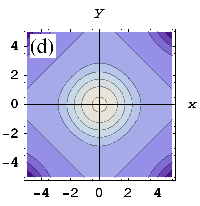}
\caption{$\mathbf{Fig.5}$ (Color online)  Plots for the cold atoms $^{87}Rb$ localization: contour plots of $\chi^{''}$ as a function of (x,y) in dependence on the incoherence pump field $\Gamma$. All the parameters from (a) to (d) are the same as in Figs.4 from (a) to (d), respectively.}
\end{figure}\label{Fig.5}

\section{CONCLUSION}

To sum up, our scheme achieved a precise localization for the cold atoms $^{87}Rb$ via the simulation
of three-level atomic system, in which the standing-wave field couples the transition $|5S_{3/2},F$=$2\rangle$-$|5S_{1/2},F$=$1\rangle$
, and the transition $|5S_{3/2},F$=$2\rangle$-$|5S_{1/2}$,$F$=$2\rangle$ is driven by the weak probe field and the incoherent pump field.
Within 0.5 wavelength domain, the increasing single absorption peak maxima for the cold atoms $^{87}Rb$ localization is obtained by the
manipulated intensities of SGC, and the incoherent pump field incurs the single peak maxima with shrinking diameters.
These results demonstrate a flexible parameter manipulation on the cold atoms $^{87}Rb$ localization, and
we anticipate this would come true in future
\begin{acknowledgments}
This work is Supported by the National Natural Science Foundation of China ( Grant Nos. 61205205 and 6156508508 ),
the General Program of Yunnan Provincial Research Foundation of Basic Research for application, China ( Grant No. 2016FB009 )
and the Foundation for Personnel training  projects of Yunnan Province, China ( Grant No. KKSY201207068 ).
\end{acknowledgments}


\end{document}